# What is the ground-state structure of intermediate-sized carbon clusters?

Ming Yu, Indira Chaudhuri, C. Leahy, C. S. Jayanthi, and S. Y. Wu
*University of Louisville, Department of Physics, Louisville, KY 40292*

(Dated: June. 28, 2008)

## ABSTRACT

A comprehensive study on the relative structural stability of various nanostructures of carbon clusters (including fullerenes, cages, onions, icosahedral clusters, bucky-diamond clusters, spherically bulk terminated clusters, and clusters with faceted termination) in the range of $d < 5$ *nm* has been carried out using a semi-empirical method based on a self-consistent and environment-dependent/linear combination of atomic orbital (SCED-LCAO) Hamiltonian. It was found that among these nanostructures with the same diameter, fullerenes are still the most stable structure, in contrast to the icosahedral cluster being the ground state structure for a series of discrete *n* values for other tetravalent clusters. The transformations from a bucky-diamond structure to an onion structure, or to a cage structure, or from an onion structure to a cage structure have been observed using a finite temperature molecular dynamics scheme based on the SCED-LCAO Hamiltonian. It was also found that the size-dependence of the HOMO-LUMO gap of fullerene shows an oscillation as a function of its diameter (*d*). Such oscillation is associated with the symmetry of the fullerene, and the magnitude of oscillation appears to decrease as its size increases.

I. Introduction

In general, a tetravalent atom such as a Si atom or a Ge atom prefers to form exclusively $sp^3$ bonding in a cluster. Therefore, the stable configuration of a cluster composed of tetravalent atoms usually assumes the form with a tetrahedral network in the interior and a well reconstructed surface in the exterior. Recently, Zhao et al. [1] have found that for Si clusters with the diameter less than 5 nm, clusters with the icosahedral structure are the most stable ones. In contrast to the Si or the Ge atom, elemental carbon could exist in three bonding states corresponding to $sp$, $sp^2$, and $sp^3$ hybridization of the atomic orbitals. Therefore, the stable configuration of a carbon cluster can assume the form of a monocyclic structure (e.g. benzene), or a closed-shell structure (e.g. fullerenes or onions), or a nano-diamond structure, or even a combination of the diamond structure and the graphitic-like structure such as bucky-diamond clusters which were discovered by Lawrence Livermore group [2]. Thus, it is important to investigate the relative structural stability among these clusters. Specifically, what is the relative stability among the carbon clusters if the newly identified icosahedral structure as those found in Si clusters and the bucky-diamond structure are included? It is also important to study how the bonding nature of carbon atoms in a cluster affects the relative stability of the cluster.

Recently, simulation results on some carbon clusters have been generated by several groups. J-Y. Raty and G. Galli *et al*. [2] have studied the relative stability of nanodiamonds as a function of size and of the surface hydrogen coverage using DFT based calculation. Their results show that as the size of nanodiamond is reduced to about 3 nm, it is energetically more favorable for this material to have bare, reconstructed surfaces than hydrogenated surfaces. This inability to retain hydrogen at the surface may then prevent the growth of larger grains. Barnard and colleagues [3] included fullerenes and onions in the analysis of the relative stability of diamond and graphite at the nanoscale using a model based on the atomic heat of formation and defined a size region of the coexistence of bucky diamond with nanodiamond and fullerene phases. As the system size increases the most stable structure of carbon clusters at the nanoscale changes from the onion (~1.4 nm < $d$ < ~1.7 nm) — to the coexistence of onion and bucky diamond (~1.7 nm < $d$ < ~2.0 nm) — to the coexistence of bucky diamond and nanodiamond (~2.0 nm < $d$ < ~2.2 nm). Although a large number of simulation results have been generated for carbon clusters, a definite conclusion on the relative stability of these clusters is still lacking because these results were obtained by different methods. Therefore, it is desirable that the simulations be carried out using the same computational approach (of an *ab initio* caliber) so that a quantitative comparison of energetics is possible. This is the purpose of this paper. We have preformed a comprehensive study on the relative stability of the intermediate size of carbon clusters (with the diameter $d$ < 5 nm) including various classes of carbon families. This is not a simple task due to a large numbers of the possible configurations in these families. Even though the size of the clusters considered is within a few nm range, it still requires a tremendous of amount of computational power to treat all these clusters in a reasonable time frame. In order to accomplish objectives such as this, we have recently developed a semi-empirical method based on a self-consistent (SE) and environment-dependent (ED) multi-center Hamiltonian in the framework of linear combination atomic orbitals (LCAO) [4]. In this approach, the SCED-LCAO Hamiltonian is constructed with a careful modeling and optimization of parameters for electron-electron correlations and multi-center interactions. It has been shown in our previous paper [4] that the SCED-LCAO Hamiltonian constructed for silicon has the *reliability* and the *transferability* for quantum mechanics-based simulations of a wide range of silicon-based systems and is able to capture the effect of screening by electrons for different condensed phases, including 0-, 1-, 2-, and 3-dimensional systems. Thus, it appears that the SCED-LCAO indeed possesses the predictive power and can be used to study the nano-scale systems with both the size and the simulation time beyond the scope of density functional theory (DFT)-based methods.

II. The Method

Before studying the structural stability among various forms of carbon clusters, we first need to construct the SCED-LCAO Hamiltonian for carbon. Using the fitting procedure for the parameters of the SCED-LCAO Hamiltonian [4], we first constructed a database of a wide range of physical properties of carbon-based systems using DFT-based methods [5]. The physical properties in the database include the bond length and the binding energy of the carbon clusters of small sizes, and cohesive energy and band structure of some bulk phases. The resulting SCED-LCAO parameters for carbon are listed in Table 1. In Table 2, the bond length and the binding energy of both the stable and some of the metastable phases of the carbon clusters of up to six atoms, obtained by the SCED-LCAO Hamiltonian and an *ab initio* method [5] respectively, are shown. It can be seen that the results by the two methods agree very well. The various phase diagrams of bulk carbon obtained from the SCED-LCAO approach are shown in Figure 1. They agree excellently with the DFT calculations [6]. These results demonstrate that the SCED-LCAO Hamiltonian can provide the accuracy, the reliability and the transferability for carbon-based systems. We also carried out an energy minimization of a $C_{147}$ cluster initially constructed from the diamond structure by the spherical bulk truncation, using both the SCED-LCAO Hamiltonian and a two-center tight binding Hamiltonian [7]. We found that the SCED-LCAO Hamiltonian successfully reproduced the bucky-diamond structure (a diamond-like core of 35 atoms connected to a fullerene shell of 112 atoms) obtained by a DFT-based method [2]. However, the two-center TB scheme failed to reproduce the bucky-diamond structure using the same dynamical relaxation process. This demonstrated further the robustness of the SCED-LCAO Hamiltonian for carbon.

III. Results and Discussion

Using the SCED-LCAO Hamiltonian for carbon listed in Table 1, we have performed a comprehensive study on the structural stability of families of stable carbon clusters that have been observed, or could likely be synthesized. In our study, we considered various types of initial configurations of carbon clusters corresponding to the diamond-like structure with either spherical or facetted bulk truncation, the icosahedral structure, and the close-shell structure including the fullerene (with only pentagons and hexagons), the cage structure (other than fullerenes, with other polygons in addition to pentagons and hexagons), and the onion structure. In the case of facetted bulk truncation, the initial configurations were constructed with no more than one dangling bond on surface atoms. For each family of carbon clusters, we considered a series of initial configurations with the size of the cluster up to about 5 nm (~ 11000 atoms). All of the initial configurations were relaxed until the force on each atom of a given system is less than 0.01 eV/Å. In the case of clusters with bulk truncation, the order-$N$ scheme [8] was applied to carbon clusters of very large sizes (with $N > 1500$).

The total energy per atom of various types of carbon clusters as a function of the total number of atoms $N$ for $N$ up to 1,500 is shown in Figure 2. The results with bulk truncated structures are denoted by full (spherical truncation) and open (facetted truncation) circles, respectively. It can be seen that in addition to the general trend of decreasing energy with respect to the total number of atoms, one can also identify some explicit local minima for the spherically truncated clusters. It turns out that some of these local minima correspond either to the bucky-diamond cluster ($C_{147}$), or bucky-diamond-like clusters ($C_{275}$ and $C_{705}$) (Ref [2]). This is really quite remarkable since it indicates that the relaxation scheme based on the SCED-LCAO Hamiltonian can lead directly from the initial configurations of spherical bulk truncation to bucky-diamond-like structures at precisely $N$=147, 275, and 705, demonstrating the robustness of the SCED-LCAO Hamiltonian. We have also carried out a detailed analysis of the relaxed structures of the clusters of spherical bulk truncation to shed light on the existence of the local minima. In Table 3, relevant quantities characterizing the relaxed structures of carbon clusters of spherical bulk truncations with completed shells are listed. It can be seen that while local minima occur at $N$ = 29, 71, 147, 275, 381, 465, 633, and 705 (see Fig. 2), only those at N = 147, 275, and 705 share the same unique property, namely, the atoms in these clusters have no more than one "dangling bond". It should be noted that for a tetravalent atom, the dangling bond associated with the atom is defined with respect to the complete $sp^3$ bonding nature with the atom having four nearest neighbors. Thus, atoms in a cluster with no more than one dangling bond indicates that atoms in this cluster either have three (one dangling bond) or four nearest neighbors (no dangling bond). Since carbon atoms can form either strong $sp^2$ bonds (with three nearest neighbors) or $sp^3$ bonds (with four nearest neighbors), a carbon cluster with its atoms having no more than one dangling bond is therefore a more stable cluster compared to its neighboring clusters with their component atoms having more than one dangling bond. Hence, for the curve of the energy per atom of spherically bulk truncated carbon clusters, the clusters with N=147, 275, or 705 are the local minima with respect to their neighbors. Furthermore, in the relaxed configuration of spherically truncated clusters with no more than one dangling bond, those atoms having four nearest neighbors (with $sp^3$ bonding) will remain in the diamond structure while those atoms with three nearest neighbors (with $sp^2$ bonding) will form fullerene-like shell, leading to the bucky-diamond or bucky-diamond-like structure.

In Figure 2, the energy per atom as a function of $N$ for the cases of icosahedral structure (full squares), the cage structure (down triangles), and the fullerene structure (up triangles) is also shown. It can be seen that for these three cases, the energy per atom as a function of $N$ also exhibits a general decreasing pattern, with the icosahedral structure having lower energy per atom compared to the bulk truncated structure, followed by the cage structure, and then the fullerene structure with the lowest energy among all families of carbon clusters.

In the inset of Fig. 2, the trend of the energy per atom as a function of $N$ exhibited by the families of carbon clusters indicates that fullerenes have the lowest energy among all families of carbon clusters for $N$ up to ~11,000 since the energy per atom of fullerenes has already approached the energy per atom of the graphite, the lowest energy among all carbon-based systems, for $N \geq 1{,}300$ while the energy per atom of bulk truncated clusters has been leveling off to about 0.3 eV/atom higher than that of the graphite for $N \approx 11{,}000$. That fullerenes have the lowest energy among all other carbon clusters with the same number of carbon atoms can be simply attributed to the fact that carbon atoms in fullerenes are bound by $sp^2$ bonds. To further delineate this situation, we plotted, in Fig. 3, the relative energy per atom with respect to the energy per atom of the graphite for three families of clusters, namely, clusters with the bucky-diamond structure, clusters of the icosahedral structure, and fullerenes, as a function of the diameter of the cluster. It should be noted here that the diameter $d$ of a cluster is defined through either (1) $4\pi \dfrac{(d/2)^3}{3} = N(\dfrac{a^3}{8})$ for bulk truncated and icosahedral clusters with $a$ being the bulk lattice constant or (2) $4\pi(d/2)^2 = N(\dfrac{a^2 \sin(\pi/3)}{2})$ for fullerenes, cages, and onions with $a$ being the graphite lattice constant. Obviously, for the same size ($d$) but different types of carbon clusters, the total number of atoms $N$ is larger in bulk-truncated or icosahedral clusters than in cage-like clusters. The behavior patterns of the energy per atom exhibited by the three families of clusters shown in Fig. 3 are all monotonically decreasing functions of $d$ as expected. The trends shown by the patterns of clusters with the bucky-diamond structure and that with the icosahedral structure indicate both patterns leveling off at about $d = 4$ nm and a likely cross-over between the two structures for $d > 4$ nm with the energies per atom for both structures above that of bulk diamond while the pattern exhibited by the fullerene structure has already approached the energy per atom of the graphite for $d \approx 4$ nm. Thus, it appears that even for $d > 4$ nm, the clusters with the fullerene structure is the most stable phase among all families of carbon clusters. We fitted the three behavior patterns to a functional form $E = E_0 + \alpha/d^\beta$, with $E_0 = 0.20$ eV/atom, $\alpha = 5.8$ eV-Å/atom, $\beta = 1$ for the bucky-diamond structure; $E_0 = 0.23$ eV/atom, $\alpha = 3.4$ eV-Å/atom, $\beta = 1$ for the icosehedral structure; and $E_0 = 0$, $\alpha = 4.0$ eV- Å$^{1.4}$/atom, $\beta = 1.4$ for fullerenes. Using the parameter sets characterizing the $d$-dependent functions for the bucky-diamond structure and the icosahedral structure respectively, a cross-over diameter between the two structures is estimated at about $d = 8.0$ nm. Although the parameter fitting for the bucky-diamond structure and that for the icosehedral structure are based only on a few data points so that the cross-over predicted at $d = 8.0$ nm may not be accurate, the existence of the cross-over between the two structures beyond 4 nm is correct since the energy per atom for bulk truncated clusters will approach that of bulk diamond while the icosahedral structure will be unstable with respect to the diamond phase as $d \rightarrow \infty$. Furthermore, this scenario of cross-over is also consistent with the cross-over between the icosahedral clusters and bulk truncated clusters observed for silicon clusters by Zhao *et al.* [1]. On the other hand, the $d$-dependence of the fullerene clearly indicates that there can be no cross-over between all the other carbon clusters. In Fig. 3, we have also shown by stars the results of the calculation of the energy per atom for the three types of clusters (the bucky-diamond, icosahedral, and fullerene structure) up to $N = 280$ ($d \leq 1.5$ nm) using the DFT-based VASP (Ref [9]). The trend exhibited by the SCED-LCAO results is seen to be consistent with the patterns of behavior of VASP results, validating the conclusion drawn from the SCED-LCAO for larger clusters.

There was an experimental observation of the phase transition between a buck-diamond-like structure and an onion-like structure [10], To shed light on this transition, we have relaxed and calculated the energy of four fullerene onions composed of two fullerene shells, namely $C_{20}@C_{150}$, $C_{60}@C_{180}$, $C_{60}@C_{240}$, and $C_{80}@C_{320}$, using the SCED-LCAO Hamiltonian. We chose these four structures because their diameters are close to the diameters of bucky-diamond clusters $C_{147}$ and $C_{275}$. From Fig. 3, it can be seen that the pattern of the energy per atom of these double-shell fullerene onions falls between that of the icosahedral clusters and that of the fullerene, i.e., the energy per atom of those double-shell fullerene onions is lower than those of bucky-diamond clusters and icosahedral clusters but still higher than that of fullerenes, of the same diameter. Therefore it is energetically very likely for a bucky-diamond-like structure to undergo a transition to an onion-like structure of the same number of atoms. Hence, we carried out a finite-temperature molecular dynamics simulations based on the SCED-LCAO Hamiltonian. We found, as shown in Fig. 4, that when the bucky-diamond $C_{147}$ cluster is heated up to 2500 K and subsequently annealed to 0 K, bonds connecting the diamond core to the fullerene-like surface are first broken, leading to an extension of the outer shell. Then the inner diamond core becomes unstable and stabilizes to a cage structure to reduce the dangling bonds, forming a double-shell onion-like structure (with outer-shell having 118 atoms and the inner-shell 28 atoms). Thus, we have demonstrated the transition from a bucky-diamond structure to an onion-like structure, similar to that observed in the experiment [10]. We have also observed the transition from an onion-like structure to a cage-like structure by heating the onion-like $C_{147}$ cluster up to 3200 K. In this simulation, bonds in the inner shell are broken and all atoms in the interior move out to merge to the outer shell. Another transition observed in our study is the one from a bucky-diamond phase directly to a cage-like phase. We found that when the bucky-diamond $C_{147}$ cluster is heated to 3500 K gradually, the inner core is completely decomposed and all of the atoms in the interior have enough energy to move out to merge to the surface forming a cage structure. This result is also consistent with the scenario presented in Fig. 3.

The electronic structures of carbon clusters have also been studied. The HOMO-LUMO gaps of fullerenes calculated from the SCED-LCAO approach are shown in figure 4. The DFT results are also shown for comparison (Ref [9]). It can be seen that they agree with each other very well. We found that the size-dependence of HOMO-LUMO gap of fullerenes shows an oscillation as a function of their diameter (*d*) for *d* < 5 nm, with its magnitude ranging from close to 2 eV to only about 100 meV. Such oscillation is associated with the symmetry of the fullerenes. The magnitude of the oscillation appears to decrease as its size increases as expected since the properties of a fullerene will approach those of a graphene sheet.

IV. Conclusion

In the present study, we have conducted a comprehensive study of the energetics of carbon clusters that have been observed or likely to be synthesized, including bulk truncated clusters, bucky-diamonds, icosahedral clusters, clusters with cage structures, clusters with onion structures, and fullerenes, using the SCED-LCAO Hamiltonian constructed for carbon. We have established that among these stable carbon clusters of the same diameter, fullerenes have the lowest energy because atoms in fullerenes have only $sp^2$ bonding. We have investigated the experimentally observed phase transition between a bucky-diamond like structure and an onion-like structure. We have shown that heating and subsequently cooling the bucky-diamond $C_{147}$ will lead it to undergo a transition to a double-shell onion-like structure using a finite temperature molecular dynamics scheme based on the SCED-LCAO Hamiltonian. In the simulation, we found that further heating could lead to the transition from the double-shell onion-like structure to a cage structure. We also found that heating the bucky-diamond at even more elevated temperature could induce a direct transition from the bucky-diamond to the cage structure. This result may prove to be a viable pathway for the synthesis of large cage structures. We also found that the size-dependence of HOMO-LUMO gaps of fullerenes shows an oscillation as a function of their diameter (*d*). Such oscillation is associated with the symmetry of the fullerenes and the magnitude of oscillation appears to decrease as its size increases.

Table 1: Parameters of the SCED-LCAO Hamiltonian for carbon.

| | | | |
|---|---|---|---|
| $\varepsilon_s$ (eV) | -17.360 | $\alpha_{ss\sigma}$ (Å$^{-1}$) | 2.153 |
| $\varepsilon_p$ (eV) | -8.329 | $d_{ss\sigma}$ (Å) | 0.629 |
| $\varepsilon'_s$ (eV) | -35.712 | $B_{sp\sigma}$ (Å$^{-1}$) | -0.777 |
| $\varepsilon'_p$ (eV) | -22.153 | $\alpha_{sp\sigma}$ (Å$^{-1}$) | 2.013 |
| $\alpha_K$ (Å$^{-1}$) | -0.0329 | $d_{sp\sigma}$ (Å) | 0.782 |
| $U$ (eV) | 14.896 | $B_{pp\sigma}$ (Å$^{-1}$) | -1.895 |
| $B_Z$ (Å$^{-1}$) | 1.475 | $\alpha_{pp\sigma}$ (Å$^{-1}$) | 1.881 |
| $A_N$ (eV) | -2.539 | $d_{pp\sigma}$ (Å) | 0.377 |
| $B_N$ (Å$^{-1}$) | -1.798 | $B_{pp\pi}$ (Å$^{-1}$) | 0.236 |
| $\alpha_N$ (Å$^{-1}$) | 3.115 | $\alpha_{pp\pi}$ (Å$^{-1}$) | 2.255 |
| $d_N$ (Å) | 0.800 | $d_{pp\pi}$ (Å) | 0.547 |
| $B_{ss\sigma}$ (Å$^{-1}$) | 0.228 | $R_{cut}$ (Å) | 4.0 |

Table 2: Comparisons of bond lengths (Å) and binding energies (eV) for different relaxed geometries of carbon clusters obtained using the SCED-LCAO Hamiltonian and *ab initio* calculations [5].

| Cluster | Symmetry | Present work | *ab initio* values[a] |
|---------|----------|--------------|------------------------|
| $C_2$   | $D_{ih}$ | 1.293        | 1.244                  |
|         |          | -5.228       | -4.527                 |
| $C_3$   | $D_{ih}$ | 1.329        | 1.287                  |
|         |          | -6.588       | -6.586                 |
| $C_3$   | $C_{2v}$ | 1.326        | 1.256                  |
|         |          | 1.515        | 1.459                  |
|         |          | -5.988       | -6.225                 |
| $C_4$   | $D_{2h}$ | 1.488        | 1.439                  |
|         |          | -6.698       | -6.746                 |
| $C_4$   | $D_{ih}$ | 1.324        | 1.288                  |
|         |          | 1.361        | 1.306                  |
|         |          | -6.520       | -6.620                 |
| $C_4$   | $D_{2d}$ | 1.382        | 1.316                  |
|         |          | 1.554        | 1.555                  |
|         |          | -5.631       | -5.566                 |
| $C_4$   | $T_d$    | 1.577        | 1.621                  |
|         |          | -5.510       | -4.830                 |
| $C_5$   | $D_{ih}$ | 1.325        | 1.277                  |
|         |          | 1.341        | 1.282                  |
|         |          | -7.124       | -7.319                 |
| $C_5$   | $D_{3h}$ | 1.487        | 1.488                  |
|         |          | 2.113        | 2.013                  |
|         |          | -6.917       | -6.578                 |
| $C_5$   | $C_{4v}$ | 1.495        | 1.443                  |
|         |          | 1.607        | 1.668                  |
|         |          | -6.547       | -6.242                 |
| $C_5$   | $T_d$    | 1.409        | 1.417                  |
|         |          | 2.301        | 2.314                  |
|         |          | -5.521       | -5.100                 |
| $C_6$   | $D_{3h}$ | 1.909        | 1.823                  |
|         |          | 2.090        | 1.864                  |
|         |          | -6.995       | -7.443                 |

| $C_6$ | $D_{6h}$ | 1.349 | 1.298 |
| | | -6.985 | -7.297 |
| $C_6$ | $D_{ih}$ | 1.332 | 1.270 |
| | | 1.329 | 1.285 |
| | | 1.355 | 1.294 |
| | | -7.054 | -7.291 |
| $C_6$ | $D_{4h}$ | 1.519 | 1.536 |
| | | 1.824 | 1.790 |
| | | -6.909 | -6.467 |
| $C_6$ | $D_{5v}$ | 1.406 | 1.354 |
| | | 1.689 | 1.698 |
| | | -6.158 | -6.254 |

a: Ref [5].

Table 3. Analysis of dangling bonds associated with carbon clusters with spherical bulk truncation, where $N_{atom}^{n\ dangling\ bonds}$ denotes the number of atoms with $n$ dangling bonds

| $C_n$ | $N_{atom}^{1\ dangling\ bond}$ | $N_{atom}^{2\ dangling\ bonds}$ | $N_{atom}^{3\ dangling\ bonds}$ |
|---|---|---|---|
| $C_{17}$ | 12 | | |
| $C_{29}$ | 23 | | |
| $C_{35}$ | 16 | 6 | |
| $C_{47}$ | 24 | 6 | |
| $C_{71}$ | 22 | 5 | |
| $C_{87}$ | 30 | 17 | |
| $C_{99}$ | 34 | 15 | 1 |
| $C_{123}$ | 56 | 12 | |
| $C_{147}$ | 115 | | |
| $C_{159}$ | 100 | | |
| $C_{167}$ | 64 | 12 | 4 |
| $C_{191}$ | 59 | 10 | 2 |
| $C_{239}$ | 60 | 17 | 1 |
| $C_{275}$ | 144 | | |
| $C_{281}$ | 121 | 5 | |
| $C_{293}$ | 148 | 6 | |
| $C_{329}$ | 76 | 30 | 12 |
| $C_{357}$ | 95 | 24 | |
| $C_{381}$ | 132 | 18 | |
| $C_{417}$ | 117 | 32 | |
| $C_{441}$ | 128 | 24 | |
| $C_{465}$ | 88 | 12 | |
| $C_{489}$ | 136 | 36 | |
| $C_{525}$ | 140 | 36 | 8 |
| $C_{597}$ | 156 | 36 | |
| $C_{633}$ | 106 | 24 | |
| $C_{657}$ | 192 | 12 | |
| $C_{705}$ | 228 | | |
| $C_{729}$ | 240 | 6 | |

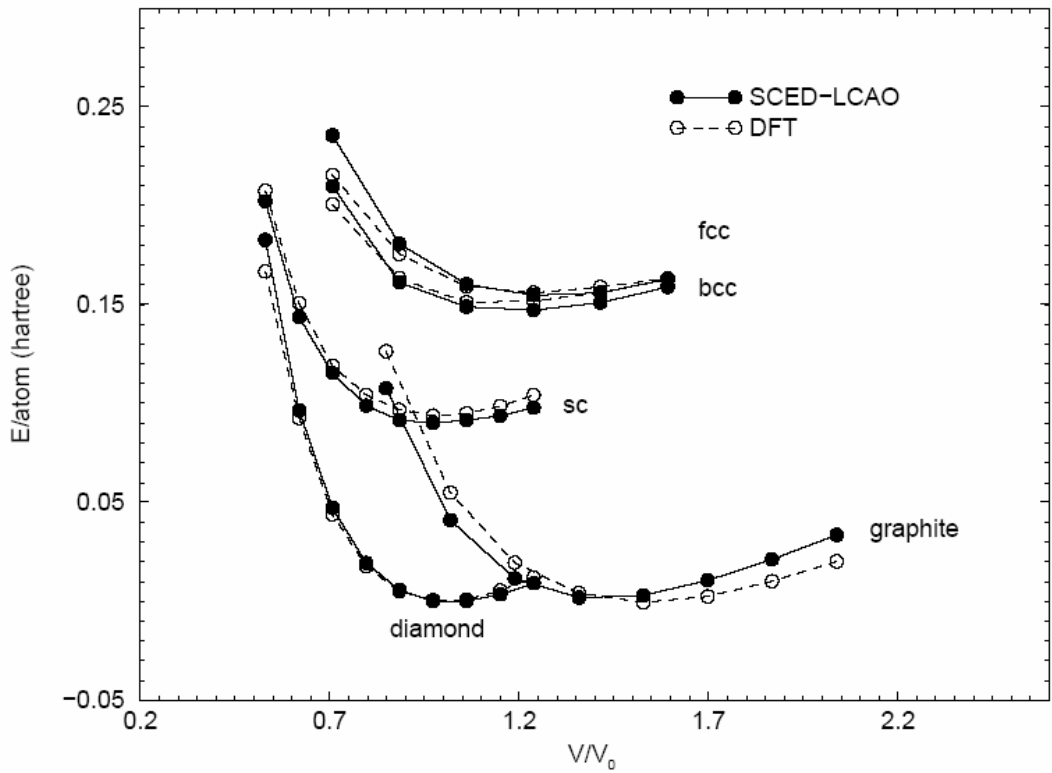

Figure 1. The binding energy per atom versus relative atomic volume (equilibrium atomic volume for diamond) for the diamond, the graphite, the simple cubic (sc), the body centered cubic (bcc), and the face centered cubic (fcc) phases of carbon, obtained using the present SCED-LCAO scheme (solid curves) and compared with the results obtained by a DFT-LDA calculation [6] (dotted curves).

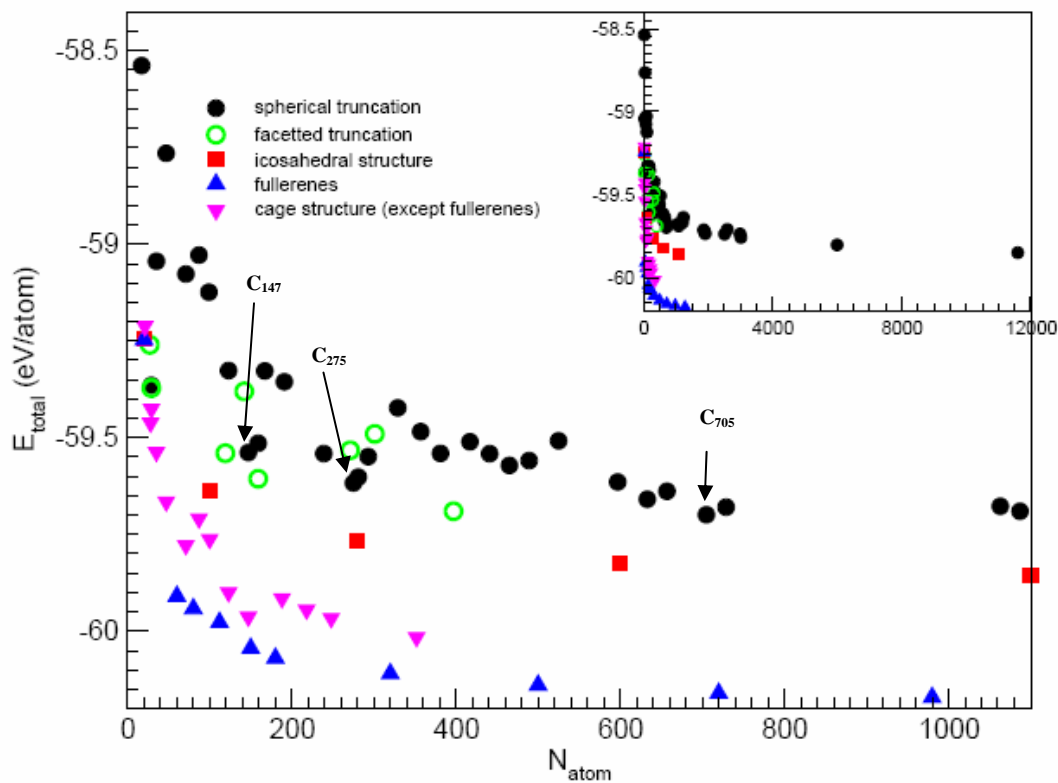

Figure 2. The total energy per atom as a function of the number of atoms $N_{atom}$ for carbon clusters with the spherical bulk truncation (full circles), the facetted bulk truncation (open circles), and the icosahedral structure (full squares), as well as fullerenes (up triangles) and cages (clusters with the cage structure except fullerenes) (down triangles). The insert shows the results including clusters of larger size with spherical bulk truncation up to 11603 atoms.

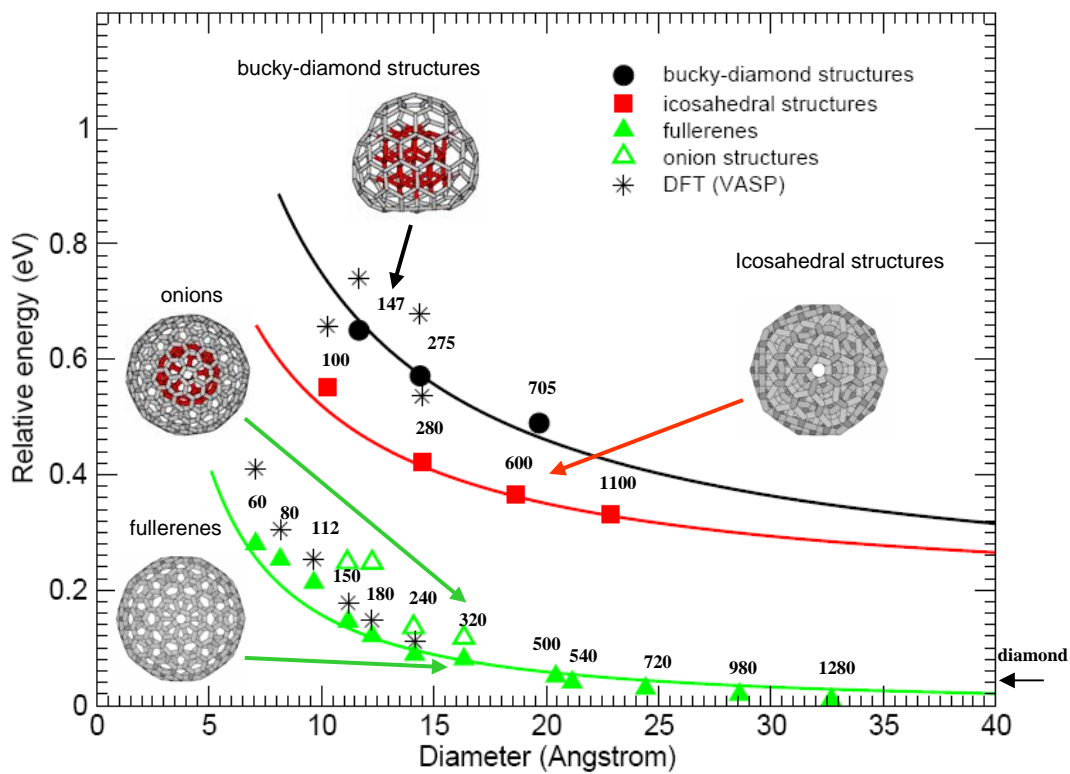

Figure 3. Relative energies per atom with respect to the energy per atom of the graphite as a function of the diameter (see in text) for three families of clusters: bucky-diamond clusters (full circles), icosahedral clusters (full squares), and the fullerenes (full triangles). The relative energy per atom of some onion structures are also shown by open triangles. The DFT (VASP) [9] results are presented by stars for comparison. The curves are presented as guidelines to the relative energy of the bucky-diamond cluster, the icosahedral cluster, and the fullerene, respectively. The arrow indicates the relative energy per atom of diamond with respect to the energy per atom of the graphite.

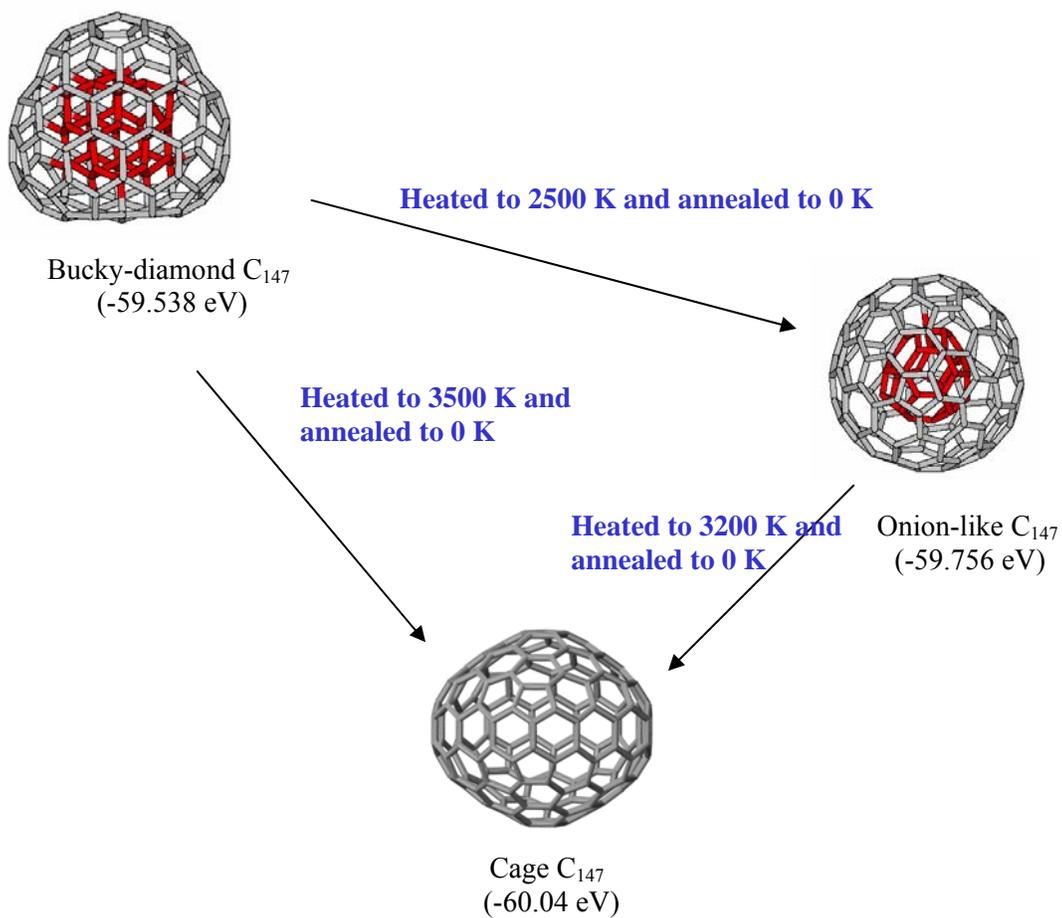

Figure 4. A schematic illustration of the phase transition of $C_{147}$ from a bucky-diamond structure (top-left) to an onion-like structure (middle-right), or from a bucky-diamond structure to a cage structure (bottom), or from an onion structure to a cage structure, respectively. The corresponding transition conditions are indicated accordingly. The total energy per atom for each phase is indicated in the parentheses.

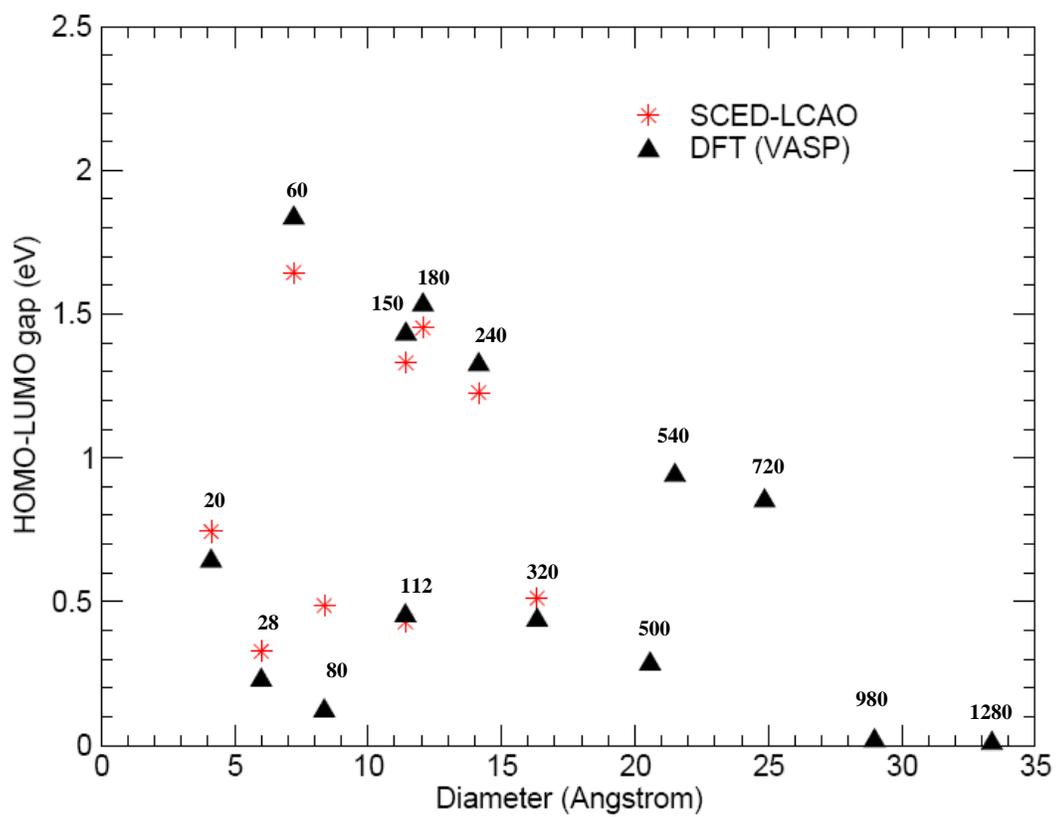

Figure 5. The HOMO-LUMO gap of fullerenes as a function of the diameter. The DFT results using VASP [9] are also shown (stars) for comparison.